\documentclass[superscriptaddress,showpacs,preprintnumbers,amsmath,amssymb,twocolumn,prl]{revtex4}

\usepackage{graphicx}
\usepackage{dcolumn}
\usepackage{bm}
\usepackage{color}
\usepackage{braket}

\newcommand{\comment}[1]{}
\def\be{\begin{equation}}
\def\ee{\end{equation}}
\def\bea{\begin{eqnarray}}
\def\eea{\end{eqnarray}}

\begin{document}

\title{Quantum beats in the polarization of the spin-dependent photon echo from donor-bound excitons in CdTe/(Cd,Mg)Te quantum wells}

\author{S.~V.~Poltavtsev}
\email{sergei.poltavtcev@tu-dortmund.de}
\affiliation{Experimentelle Physik 2, Technische Universit\"at Dortmund, 44221 Dortmund, Germany}
\affiliation{Spin Optics Laboratory, St.~Petersburg State University, 198504 St.~Petersburg, Russia}
\author{I.~A.~Yugova}
\affiliation{Spin Optics Laboratory, St.~Petersburg State University, 198504 St.~Petersburg, Russia}
\affiliation{Physics Faculty, St. Petersburg State University, 199034, St. Petersburg, Russia}
\author{Ya.~A.~Babenko}
\affiliation{Spin Optics Laboratory, St.~Petersburg State University, 198504 St.~Petersburg, Russia}
\affiliation{Physics Faculty, St. Petersburg State University, 199034, St. Petersburg, Russia}
\author{I.~A.~Akimov}
\affiliation{Experimentelle Physik 2, Technische Universit\"at Dortmund, 44221 Dortmund, Germany}
\affiliation{Ioffe Institute, Russian Academy of Sciences, 194021 St.~Petersburg, Russia}
\author{D.~R.~Yakovlev}
\affiliation{Experimentelle Physik 2, Technische Universit\"at Dortmund, 44221 Dortmund, Germany}
\affiliation{Ioffe Institute, Russian Academy of Sciences, 194021 St.~Petersburg, Russia}
\author{G.~Karczewski}
\affiliation{Institute of Physics, Polish Academy of Sciences, PL-02668 Warsaw, Poland}
\author{S.~Chusnutdinow}
\affiliation{Institute of Physics, Polish Academy of Sciences, PL-02668 Warsaw, Poland}
\author{T.~Wojtowicz}
\affiliation{International Research Centre MagTop, Institute of Physics, Polish Academy of Sciences, PL-02668 Warsaw, Poland}
\author{M.~Bayer}
\affiliation{Experimentelle Physik 2, Technische Universit\"at Dortmund, 44221 Dortmund, Germany}
\affiliation{Ioffe Institute, Russian Academy of Sciences, 194021 St.~Petersburg, Russia}

\date{\today}

\begin{abstract}
We study the quantum beats in the polarization of the photon echo from donor-bound exciton ensembles in semiconductor quantum wells. To induce these quantum beats, a sequence composed of a circularly polarized and a linearly polarized picosecond laser pulse in combination with an external transverse magnetic field is used. This results in an oscillatory behavior of the photon echo amplitude, detected in the $\sigma^+$ and $\sigma^-$ circular polarizations, occurring with opposite phases relative to each other. The beating frequency is the sum of the Larmor frequencies of the resident electron and the heavy hole when the second pulse is polarized along the magnetic field. The beating frequency is, on the other hand, the difference of these Larmor frequencies when the second pulse is polarized orthogonal to the magnetic field. The measurement of both beating frequencies serves as a method to determine precisely the in-plane hole $g$ factor, including its sign. We apply this technique to observe the quantum beats in the polarization of the photon echo from the donor-bound excitons in a 20-nm-thick CdTe/Cd$_{0.76}$Mg$_{0.24}$Te quantum well. From these quantum beats we obtain the in-plane heavy hole $g$ factor $g_h=-0.143\pm0.005$.
\end{abstract}



\maketitle

Quantum beats are a phenomenon due to resonant coherent excitation of (at least) two discrete quantum mechanical states with different energies, leading to a superposition state. Quantum beats can be manifested by oscillations in the coherent optical response of the system due to interference of the excited polarizations, where the oscillation frequency corresponds to the energy difference between the levels \cite{ScullyBook}. A typical example are the oscillations observed in resonance fluorescence or other coherent spectroscopy techniques from excitons in semiconductors, which can be represented by V-type energy level arrangements with a common crystal ground state that is optically coupled to two split excited exciton states \cite{KochSSC1993, Feuerbacher1990, LeoPRB1991, StolzPRL1991, PoelOC1990, Gourdon1992, Flissikowski2001, Gilliot1999, Moody2014, Singh2014, Hao2016, Carmel1993, Li2004, Trifonov2015}. Quantum beats on excitons in semiconductors and their nanostructures have been observed for a large variety of  excited states corresponding to the beating between, e.g.,  heavy- and light-hole excitons \cite{KochSSC1993, Feuerbacher1990, LeoPRB1991}, the exiton states of a fine structure doublet with different spin configurations \cite{StolzPRL1991, PoelOC1990, Gourdon1992, Flissikowski2001}, as well as  neutral and charged excitons \cite{Gilliot1999, Moody2014, Singh2014, Hao2016}. Application of a magnetic field can be used to split (quasi-degenerate) excitonic states by the Zeeman effect and to observe the corresponding quantum beats in the polarized optical response \cite{Carmel1993, StolzPRL1991, Li2004}. In this case, the splitting of the optically active exciton states, having opposite angular momentum projections $\pm$1 onto the quantization axis along which also the magnetic field is applied, leads to quantum beats in the polarization rather than the intensity of the emitted light. The period of the oscillations corresponds to the Zeeman splitting between the spin levels and can be used for evaluation of the $g$ factors of the excitons. Another advantage of the Zeeman effect induced quantum beats is given by the possibility to tune and control the beating frequency by the magnetic field. 

\begin{figure}[t]
	\includegraphics[width=\linewidth]{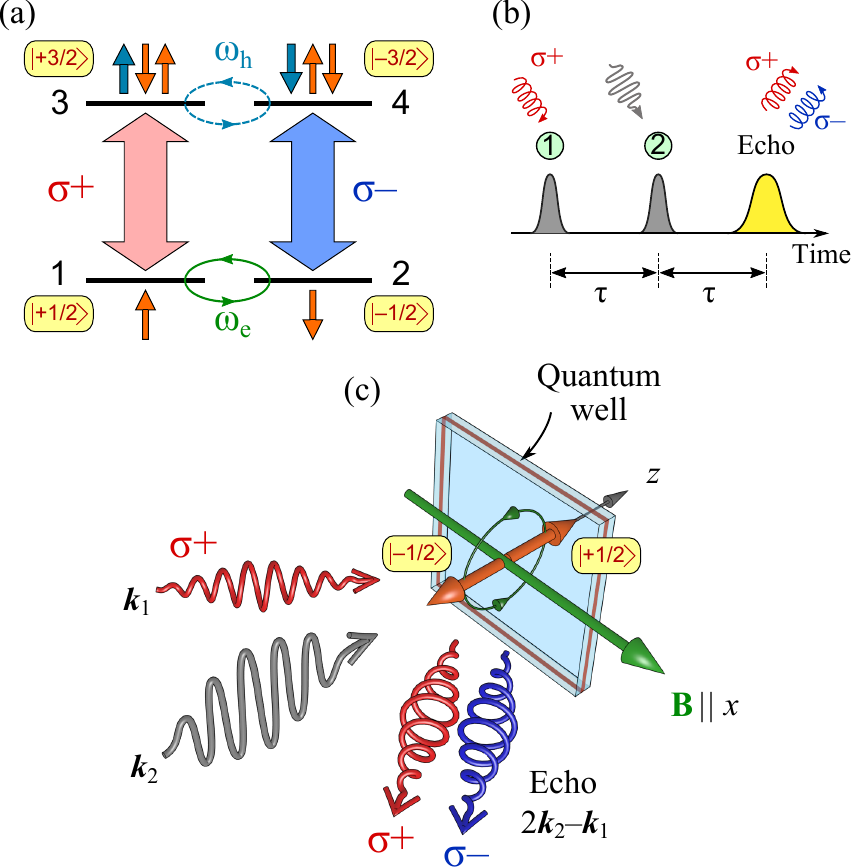}
	\caption{Photon echo from D$^0$X: (a) Energy scheme of D$^0$X. Small blue and orange arrows indicate the hole and electron spin orientations along $z$ axis, respectively. (b) Timing of PE experiment. (c) Geometry of the experiment: Excitation with $\sigma^+$ (${\bf k}_1$) and linearly polarized (${\bf k}_2$) pulses results in $\sigma^+$ or $\sigma^-$ PE ($2{\bf k}_2-{\bf k}_1$) depending on delay $\tau$ and magnetic field $B$. Circles with arrows in (a) and (c) indicate the mixing of states by the magnetic field.}
	\label{schemes}
\end{figure}

Along with neutral excitons which comprise an electron-hole pair, excitonic complexes such as donor-bound excitons and three-particle charged excitons (trions) can exist in quantum well (QW) and quantum dot structures \cite{Cox1993}. Currently, these excitonic complexes attract attention for application in spintronics since they can be used as a pathway to optically control the spin state of resident carriers in semiconductor nanostructures \cite{DyakonovBook, DeGreve2013, Linpeng2018, Gao2015}. The energy level structure of these complexes is different from the V-type energy level arrangement. The ground and lowest optically excited states are each represented by a Kramers doublet at zero magnetic field. Each of the two ground states is optically coupled to one of the excited states, and the two optical transitions have opposite circular polarizations, as shown by the two arms in Fig.~\ref{schemes}(a). A magnetic field in Faraday geometry splits the degeneracy of the doublets, but does not introduce a coupling between the two arms and, therefore, no quantum beats are observed. Here, it should be noted that most of the experiments on the coherent optical response of excitonic complexes (e.g. resonant fluorescence and four-wave mixing) were performed using the Faraday geometry. Using the Voigt geometry, the magnetic field leads to doubly coupled $\Lambda$ energy schemes \cite{Salewski2017, Gao2015, Wu2007}. Four-wave mixing experiments on trions in Voigt geometry were performed only recently. \cite{LangerPRL, LangerNature, Salewski2017, KosarevPRB2019} There quantum beats at the Larmor precession frequency were observed for the intensity of the photon echo. However, no polarization quantum beats were recorded in this system yet.

In what follows we will demonstrate that quantum beats can be induced in the polarization of the photon echo (PE) generated by donor-bound excitons (D$^0$X) by applying a sequence of circularly and linearly polarized pulses in the presence of a transverse magnetic field. These quantum beats carry information about the Larmor precession of both the resident electron and the heavy hole in D$^0$X. Therefore, the quantum beats in the PE polarization represent a tool for measuring both the in-plane electron and hole $g$ factors.

In more detail, we consider the PE generated by an ensemble of D$^0$X in a QW subject to a transverse magnetic field. Excitation by two short laser pulses separated by a time interval $\tau$ results in PE emission occurring at time $\tau$ after the second pulse, as shown in Fig.~\ref{schemes}(b). 

The D$^0$X can be represented by a four-level system as displayed in Fig.~\ref{schemes}(a). The two ground states have electrons with  total angular momentum projections $\ket{\pm1/2}$ on the $z$ axis parallel to the structure growth direction. The two excited levels correspond to the D$^0$X states with total angular momentum projections $\ket{\pm3/2}$, associated with the heavy hole spin. The left and right arms of this system can be independently addressed by circularly polarized light ($\sigma^+$ and $\sigma^-$) as dictated by the optical selection rules.

The external magnetic field is applied in the Voigt geometry (${\bf B} || x$), as illustrated in Fig.~\ref{schemes}(c). Since the basis states have spin projections perpendicular to the magnetic field axis, the ground states $\ket{\pm1/2}$ are degenerate and mixed. The electron spins precess about the $\bf B$ direction at the Larmor frequency $\omega_e=|g_e|\mu_BB/\hbar$, where $g_e$ is the in-plane electron $g$ factor, $\mu_B$ is the Bohr magneton and $\hbar$ is the Planck constant. Similarly, the two degenerate D$^0$X states with $\ket{\pm3/2}$ are mixed and the hole spins precess at the Larmor frequency $\omega_h=|g_h|\mu_BB/\hbar$, where $g_h$ is the in-plane heavy hole $g$ factor.

\begin{figure}[t]
	\includegraphics[width=\linewidth]{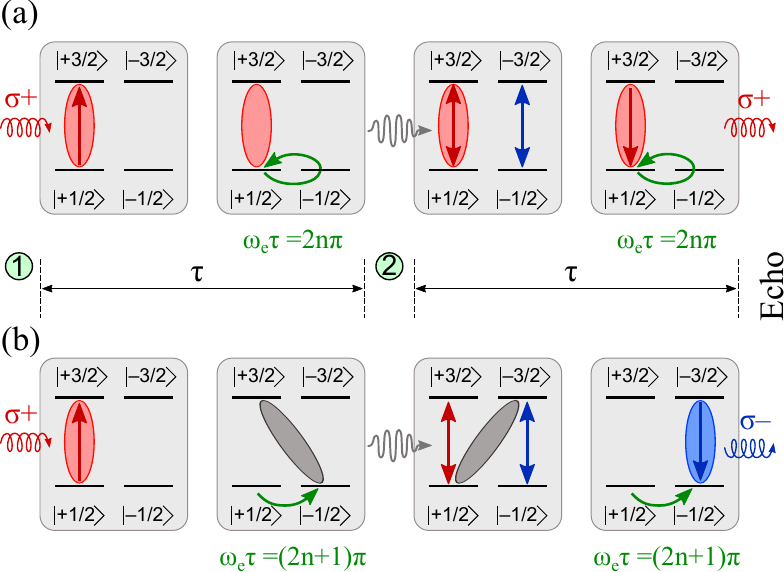}
	\caption{Temporal evolution of D$^0$X energy diagram. The hole spin precession is neglected ($g_h=0$). The first pulse is $\sigma^+$ polarized. (a) The second, linearly polarized pulse arrives after an integer number $n$ of electron spin precession periods, $\omega_e\tau=2n\pi$. The PE is emitted in the $\sigma^+$ polarization. (b) The second pulse arrives after an odd number of electron spin precession half-periods, $\omega_e\tau=(2n+1)\pi$. The PE polarization helicity is inverted to $\sigma^-$.}
	\label{invertor}
\end{figure}

The emergence of oscillations in the PE amplitude can be understood with the help of Fig.~\ref{invertor}. For simplicity, we neglect here the hole spin precession ($g_h=0$), any dispersion of the in-plane electron $g$ factor ($\Delta g_e=0$), and also relaxation processes. We assume that the pulse duration $\tau_p$ is short compared to the Larmor precession period and the delay of the second pulse: $\tau_p \ll 2\pi/\omega_e$ and $\tau_p \ll \tau$.

By the excitation with the first $\sigma^+$ polarized pulse, the ensemble of coherent superposition states on the left-hand side of the energy scheme corresponding to the $(\ket{+1/2}$,$\ket{+3/2})$ states is created. If the electron spins perform an integer number of full revolutions about the magnetic field until the second pulse arrival, then the coherent ensemble has returned to the left-hand side of the scheme, as if no magnetic field was applied [See Fig.~\ref{invertor}(a)]. The linearly polarized second pulse inverts the populations of the ground and excited states and starts the ensemble rephasing. This results in a $\sigma^+$ polarized PE emitted after the same number of electron spin revolutions as before the second pulse. When the second pulse arrives after an odd number of Larmor precession half-periods, it transfers the coherent ensemble from the $(\ket{-1/2}$,$\ket{+3/2})$ to the $(\ket{+1/2}$,$\ket{-3/2})$ superposition, as shown in Fig.~\ref{invertor}(b). As a result, a $\sigma^-$ polarized PE is emitted. Variation of either the magnetic field strength $B$ at $\tau$=const or of the delay $\tau$ at $B$=const causes a periodic alternation of the PE circular polarization between $\sigma^+$ and $\sigma^-$. 


In order to describe this effect analytically and study its consequences, we employ the spin-optical Hamiltonian, taking into account the electron and hole spin precession, in the form

\begin{align}
	\widehat{H} = \frac{\hbar}{2}
	\begin{pmatrix}
		0 & \omega_e & f^*_+e^{i\omega t} & 0 \\
		\omega_e & 0 & 0 & f^*_-e^{i\omega t} \\
		f_+e^{-i\omega t} & 0 & \omega_0 & \omega_h\\
		0 & f_-e^{-i\omega t} & \omega_h & \omega_0
	\end{pmatrix}.
\end{align}

\noindent Here, $f_\pm=-\frac{2e^{i\omega t}}{\hbar}\int d({\bm r}) E_{\sigma^{\pm}}({\bm r},t)d^3r$ are the envelopes of the circularly polarized components of the light pulse with $E_{\sigma^{\pm}}$ being the electric field amplitudes with the corresponding circular polarizations, $\omega$ is the central frequency of the light pulses, which is close to the D$^0$X optical transition frequency, $\omega_0$, and $d({\bm r})=\bra{+1/2}\hat{d}_+({\bm r})\ket{+3/2}=\bra{-1/2}\hat{d}_-({\bm r})\ket{-3/2}$ are the components of the electric dipole moment operator. The Hamiltonian is composed in correspondence with the level numbering given in Fig.~\ref{schemes}(a).

The coherent evolution of the D$^0$X ensemble under the action of the optical and external magnetic fields can be described using the optical Bloch equations. We use the rectangular approximation for the optical pulses and neglect the magnetic field during the pulse action. The in-plane $g$ factors of the election and hole are considered to be isotropic.

\begin{figure}[t]
	\includegraphics[width=\linewidth]{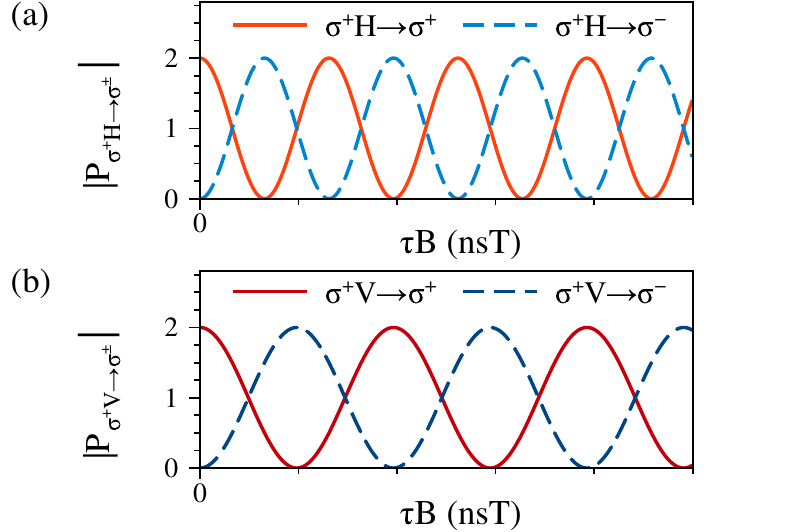}
	\caption{Dependence of the absolute value of the PE amplitude on the $\tau B$ product calculated using Eqs.~(\ref{result}). The second pulse is: (a) H polarized; (b) V polarized. Here $g_e=5g_h$ and $T_2=\infty$ is assumed.}
	\label{theory}
\end{figure}

We write down the PE amplitude detected in $\sigma^+$ or $\sigma^-$ circular polarization for two linear polarizations of the second pulse: H is the horizontal polarization parallel to the magnetic field ($||$ $\bf B$) and V is the vertical polarization orthogonal to it ($\perp\bf B$) \cite{supplement}:

\begin{gather}
	P_{\sigma^+H\rightarrow\sigma^+} \propto \big[1 + \cos(\omega_e+\omega_h)\tau\big]e^{-2\tau/T_2}, \nonumber \\
	P_{\sigma^+H\rightarrow\sigma^-} \propto \big[1 - \cos(\omega_e+\omega_h)\tau\big]e^{-2\tau/T_2}, \nonumber \\
	P_{\sigma^+V\rightarrow\sigma^+} \propto \big[1 + \cos(\omega_e-\omega_h)\tau\big]e^{-2\tau/T_2}, \nonumber \\
	P_{\sigma^+V\rightarrow\sigma^-} \propto \big[-1 + \cos(\omega_e-\omega_h)\tau\big]e^{-2\tau/T_2},
	\label{result}
\end{gather} 
\noindent where $T_2$ is the pure optical dephasing time.

Figure~\ref{theory} displays the analytically calculated absolute values of the PE amplitudes neglecting optical relaxation ($T_2=\infty$). For the H polarized second pulse, the PE oscillates between maximum amplitude and zero at the sum precession frequency, $\omega_e + \omega_h$. On the other hand, applying the V polarized second pulse results in PE oscillations at the difference precession frequency, $\omega_e - \omega_h$. The PE amplitudes detected in the $\sigma^\pm$ polarizations oscillate with opposite phases. Thus, measuring the PE oscillations for the H and V polarized second pulse, e.g. in the P$_{\sigma^+H\rightarrow\sigma^+}$ and P$_{\sigma^+V\rightarrow\sigma^+}$ configurations, allows obtaining both the in-plane electron and hole $g$ factors. The sign of $g_h$ can be also obtained, if the sign of $g_e$ is known. Equal signs of the two $g$ factors result in a higher oscillation frequency in the P$_{\sigma^+H\rightarrow\sigma^+}$ configuration than in the P$_{\sigma^+V\rightarrow\sigma^+}$ configuration, and vice versa. We note that the theory works also for trion (negative or positive), since it is described by an energy scheme similar to Fig.~\ref{schemes}(a).

Taking into account the finite dispersions of the electron and hole in-plane $g$ factors, $\Delta g_e$ and $\Delta g_h$, leading to spin dephasing, results in an exponential damping of the oscillation amplitude $\propto\exp\big[-\tau^2\mu_B^2B^2( \Delta g_e^2 + \Delta g_h^2)/2\hbar^2\big]$ for a normal distribution of the $g$ factors. Thus, applying a sufficiently large delay $\tau$ or magnetic field $B$ results in a non-oscillating PE signal that decays exponentially with the $T_2$ time constant when the delay $\tau$ is scanned.

\begin{figure}[t]
	\includegraphics[width=\linewidth]{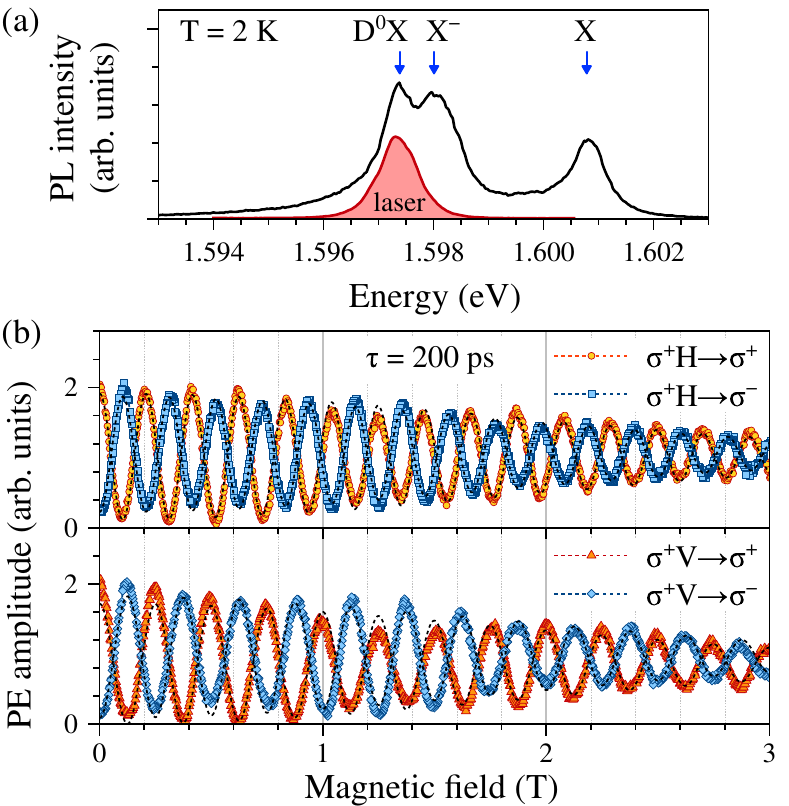}
	\caption{Summary of experimental results obtained on a 20-nm-thick CdTe/Cd$_{0.76}$Mg$_{0.24}$Te single QW: (a) PL spectrum measured at 2~K temperature (black line). Filled area indicates the laser pulse spectrum centered at 1.5973~eV. (b) Magnetic field dependences of the PE amplitude measured on D$^0$X in the $\sigma^+$H$\rightarrow\sigma^\pm$ and $\sigma^+$V$\rightarrow\sigma^\pm$ configurations. Pulse delay $\tau=200$~ps. Data are normalized to the amplitude of 2. Dotted lines are model fits.}
	\label{data}
\end{figure}

To verify experimentally these concepts, we used a 20-nm-thick CdTe/Cd$_{0.76}$Mg$_{0.24}$Te single QW (032112B). The structure was grown by molecular-beam epitaxy on a [100]-oriented GaAs substrate. The QW layer sandwiched between 100-nm-thick Cd$_{0.76}$Mg$_{0.24}$Te barriers contains a background density of donors of $n_d < 10^{10}$~cm$^{-2}$. The photoluminescence (PL) spectrum of this QW is shown in Fig.~\ref{data}(a) and exhibits three features associated with the neutral exciton (X) at 1.601~eV, the trion (X$^-$) at 1.5980~eV, and the D$^0$X at 1.5973~eV.

The sample cooled down to about 2~K was excited by a sequence of two 2.3~ps pulses with the central energy of 1.5973~eV, whose spectrum is shown in Fig.~\ref{data}(a). The pulses with the wavevectors ${\bf k}_1$ and ${\bf k}_2$ close to the sample normal were separated by the time delay $\tau=200$~ps and focused into a spot of about 250~$\mu$m in diameter. The pulse intensities were adjusted such that they correspond to about $\pi/2$ pulse area (6~pJ) based on previous studies~\cite{Poltavtsev2017}. The PE was detected in reflection geometry along the $2{\bf k}_2-{\bf k}_1$ direction, as illustrated in Fig.~\ref{schemes}(c). A reference pulse delayed by $2\tau=400$~ps with respect to the first pulse was used for an interferometric measurement of the PE amplitude. The reference pulse polarization was set to $\sigma^+$ or $\sigma^-$ to analyze the PE in the according polarization. The detected signal intensity is $I\propto |\text{Re}(E_{PE}E_{Ref}^*)|$, where $E_{PE}$ and $E_{Ref}$ are the PE and reference pulse amplitudes, respectively. The magnetic field was applied in the QW plane. More details of the experimental technique can be found in Ref. \cite{PSS2018}.

For measuring spin-dependent PE we have chosen the D$^0$X transition since it shows better coherence properties than the X$^-$ transition. These include longer optical coherence times up to $T_2=100$~ps ($\sim80$~ps for X$^-$) and more robust optical Rabi oscillations in the PE amplitude \cite{Poltavtsev2017}. As a result, we are able to apply higher pulse powers and to set substantially longer delay times $\tau$ (200~ps), avoiding strong excitation-induced dephasing.

Measurements of the oscillating spin-dependent PE from the D$^0$X ensemble in the studied QW structure are summarized in Fig.~\ref{data}(b). The oscillations are induced by varying the magnetic field strength, resulting in a sweep of the electron and hole Larmor frequencies that scale linearly with $B$. These oscillations correspond well to the theoretical model: They are opposite in phase for $\sigma^\pm$ polarized detection and exhibit different frequencies with a ratio of about 5:4 for the H and V polarized second pulse (periods of 0.208~T$^{-1}$ and 0.250~T$^{-1}$), respectively.

These data were analyzed taking into account the heavy hole $g$ factor dispersion $\Delta g_h$. We assume $\Delta g_e\leq1\%$ and thus neglect it \cite{Saeed2018}. As a result, we find the in-plane electron and hole $g$ factors to be $g_e=-1.583\pm0.005$ and $g_h=-0.143\pm0.005$, respectively. The negative sign of the electron $g$ factor is taken here following the sign-sensitive studies on bulk CdTe \cite{Oestreich1996, Sirenko1997}. The obtained heavy hole $g$ factor dispersion is $\Delta g_h\approx0.07$.

We note that scanning of the magnetic field strength instead of the pulse delay $\tau$ allows neglecting the relaxation processes such as pure optical dephasing. Moreover, the diamagnetic shift of the D$^0$X transition within 0.3~meV, which we observe at $B=3$~T, is also negligible.

The in-plane heavy hole $g$ factor might be notably anisotropic and thus sensitive to the in-plane orientation of the applied magnetic field with respect to the crystallographic axes. The reason for that is the presence of various interactions due to heavy and light hole splitting contributing to $g_h$, such as strain-induced interaction or non-Zeeman interaction due to the cubic symmetry \cite{KusrayevPRL1999, Semenov2003, Koudinov2006}. As a result, the heavy hole experiences an effective magnetic field, which may deviate from that externally applied. Thereby, the observed oscillations can be sensitive to the angle of the sample orientation around the $z$ axis. Here we employ a fixed angle of about 90$^\circ$ between the magnetic field axis and the [010] crystal direction and do not study the in-plane $g_h$ anisotropy, which is subject of future studies.

To conclude, we have demonstrated that by applying a sequence of circularly and linearly polarized laser pulses to an ensemble of donor-bound excitons in a QW and varying the strength of the applied transverse magnetic field one can induce quantum beats in the photon echo polarization. As a consequence, the circular polarization of the photon echo can be switched between $\sigma^+$ and $\sigma^-$ in a controllable way by means of varying either the magnetic field strength or the pulse delay. From the observed oscillations in the photon echo amplitude one can precisely extract both the in-plane electron and hole $g$ factors. This method can be applied in particular to systems with the hole spin weakly interacting with the magnetic field, where other optical methods such as PL spectroscopy, pump-probe Faraday rotation, or spin-flip Raman scattering are not suitable in this case.

The authors are thankful to V. L. Korenev for fruitful discussions. This research was supported by the Deutsche Forschungsgemeinschaft through the International Collaborative Research Centre TRR 160 (Project A3). S.V.P. and I.A.Y. thank the Russian Foundation for Basic Research (Project No. 19-52-12046) and the St. Petersburg State University (Project No. 11.34.2.2012 and grant ID 40847559). The research in Poland was partially supported by the Foundation for Polish Science through the IRA Programme, co-financed by the EU within SG OP and by the National Science Centre through Grants No. 2017/25/B/ST3/02966 and 2018/30/M/ST3/00276.


\end{document}